\documentclass[11pt,preprintnumbers,aps,amssymb,nofootinbib,amsmath,superscriptaddress]
{revtex4}
\usepackage{epsfig,epsf}
\usepackage{bm} 
\newcommand{\beq}{\begin{equation}}
\newcommand{\beql}[1]{\begin{equation}\label{#1}}
\newcommand{\eeq}{\end{equation}}
\newcommand{\eq}[1]{(\ref{#1})}
\newcommand{\fig}[1]{Fig.~\ref{#1}}
\renewcommand{\sec}[1]{Sec.~\ref{#1}}
\newcounter{topiccounter}
\renewcommand{\b}[1]{{\bm #1}} 

\newcommand{\as}{\alpha_s}

\newcommand{\aver}[1]{\left\langle #1 \right\rangle}
\newcommand{\jpsi}{J\mskip -2mu/\mskip -0.5mu\psi}

\newcommand{\Q}{\mathcal{Q}}

\begin{document}


\title{Gluon saturation effects on the color singlet  $\jpsi$ production\\ in high energy $dA$ and $AA$ collisions}

\author{F.~Dominguez}
\affiliation{Department of Physics, Columbia University, New York, NY 10027, USA}

\author{D.E.~Kharzeev}
\affiliation{Department of Physics and Astronomy, Stony Brook University, Stony Brook, NY 11794, USA} 
\affiliation{Department of Physics, Brookhaven National Laboratory,
Upton, NY 11973-5000, USA}

\author{E.M.~Levin}
\affiliation{HEP Department, School of Physics,
Raymond and Beverly Sackler Faculty of Exact Science,
Tel Aviv University, Tel Aviv 69978, Israel}
\affiliation{Departamento de F\'\i sica, Universidad T\'ecnica
Federico Santa Mar\'\i a, Avda. Espa\~na 1680,
Casilla 110-V,  Valparaiso, Chile}

\author{A.H.~Mueller} 
\affiliation{Department of Physics, Columbia University, New York, NY 10027, USA}

\author{K.~Tuchin}
\affiliation{Department of Physics and Astronomy, Iowa State University, Ames, IA 50011, USA}

\date{\today}


\begin{abstract}
We derive the formulae for the cross section of $\jpsi$ production in high energy $pA$ and $AA$ collisions taking into account the gluon saturation/color glass condensate effects. We then perform the numerical calculations of the corresponding nuclear modification factors and find a good agreement between our calculations and the experimental data on $\jpsi$ production in $pA$ collisions. We also observe that cold nuclear modification effects alone cannot describe the data on $\jpsi$ production in $AA$ collisions. Additional final state suppression (at RHIC) and enhancement (at LHC) mechanisms are required to explain the experimental observations.

\end{abstract}

\maketitle

\section{Introduction}

The goal of this paper is to provide an improved analysis of the gluon saturation effects on the color singlet mechanism of  $\jpsi$ production in $dA$ and $AA$ collisions at RHIC and LHC. In our recent publications \cite{Kharzeev:2005zr,Kharzeev:2008cv,Kharzeev:2008nw}   we argued that  a mechanism responsible for $\jpsi$ production in central nuclear collisions is different from the one in $pp$ collisions. This is because the symmetry properties of $\jpsi$ under the parity and charge conjugation transformations  dictate that there must be an odd number of gluons attached to the bound $c$ and  $\bar c$ quarks. At the lowest order in strong coupling $\as$ there are three gluons attached. In $pp$ collisions, two of those gluons have their external ends attached to the valence quarks of the colliding protons whereas the third one is emitted by the $c\bar c$ dipole.  On the other hand in central $pA$ collisions the parametrically enhanced contribution in the quasi-classical regime -- which is controlled by a large parameter $\as^2A^{1/3}\sim 1$ \cite{MV,Kovchegov:1999yj,Kovchegov:1996ty} --  originates from the diagrams where one of the gluons is attached to the proton's valence quark whereas the remaining two are attached to the valence quarks inside two different nucleons of the nucleus. Obviously, such contribution breaks the perturbative QCD factorization already at the leading order in $\as$. 

In \cite{Kharzeev:2005zr} we assumed that the $c\bar c$ pair propagates through the nucleus in the color octet state and becomes color singlet only after the last interaction with the nucleus. In this paper we drop this assumption by taking into account a possibility that the $c\bar c$ pair converts from the color octet to the color singlet state already inside the nucleus. In the large $N_c$ approximation further color conversions of the $c\bar c$ state are suppressed and thus can be neglected. Therefore, in this case the $c\bar c$ experiences the last inelastic interaction inside the nucleus after which it rescatters only elastically. As a result, the last inelastic interaction does not exponentiate with the rest of the scatterings and --  as we will show -- automatically selects an odd number of inelastic scatterings as required by the parity of $\jpsi$.  This is different from our approach in \cite{Kharzeev:2005zr,Kharzeev:2008cv,Kharzeev:2008nw} where we had to select the odd number of inelastic scatterings in the scattering amplitude. Additionally, we give a more accurate treatment of $\jpsi$ wave function with parameters taken from a fit to the exclusive $\jpsi$ production in deep inelastic scattering.  

Our paper is structured as follows. In \sec{sec:pA} we derive the cross section for $\jpsi$ production in $pA$ collisions; our main result is given by Eq.~\eq{pa-m}. In \sec{sec:AA} we  propose a generalization of this result to the $AA$ collisions. The derived cross section  is given by \eq{ab-2},\eq{aa-m} and satisfies the constraints imposed by the symmetry of the $\jpsi$ wave function. The results in \sec{sec:pA} and \sec{sec:AA} are derived in the quasi-classical approximation, i.e.\ assuming that the coherence length for $\jpsi$ production is much larger than the nuclear radius, but neglecting the low-$x$ evolution. In \sec{sec:evol} we derive expression for the scattering amplitude \eq{aa-m-ev}  that includes the low-$x$ evolution and thus gives a dependence on energy and rapidity. \sec{sec:numeric} is dedicated to the description of the numerical calculations performed with different models for the dipole scattering amplitudes. Our main results are exhibited in Figs.~\ref{fig:dhj},\ref{fig:kmw}. We discuss them and conclude in \sec{sec:disc}.

\section{Production of $\jpsi$ in pA collisions} \label{sec:pA}

The cross section for $J/\psi$ production in pA collisions can be written in the factorized form
\begin{equation}
\frac{d\sigma_{pA\to J/\psi X}}{d^2bdy}=x_1G(x_1,m_c^2)\frac{d\sigma_{gA\to J/\psi X}}{d^2b}.\label{factxsec}
\end{equation}
In order to set normalizations for $\frac{d\sigma_{gA\to J/\psi X}}{d^2b}$ it is convenient to compare the $gA$ scattering process in (\ref{factxsec}) with that of $\gamma A$ where there is a well developed phenomenology. Start with $\gamma$-proton scattering where
\begin{equation}
\frac{d\sigma_{\gamma p\to J/\psi p}}{dt}=\frac{1}{16\pi}\left|A_{\gamma p\to J/\psi p}\right|^2\label{xsecphoton}
\end{equation}
with
\begin{equation}
A_{\gamma p\to J/\psi p}(x,\Delta)=\int d^2b\; e^{-i\Delta\cdot b}\int_0^1dz\int\frac{d^2r}{4\pi}\left(\psi^*_{J/\psi}\psi_\gamma\right)2i\left[1-S(x,\boldsymbol r,\boldsymbol b)\right]\label{Aphoton}
\end{equation}
and $t$ is given in terms of the momentum transfer by $t=-\Delta^2$. Call
\begin{equation}
\left(\psi^*_{J/\psi}\psi_\gamma\right)=\Phi_\gamma(\boldsymbol r,z)
\end{equation}
where
\begin{equation}
\Phi_\gamma(\boldsymbol r,z)=\frac{2}{3}e\frac{N_c}{\pi}\left\{m_c^2K_0(m_cr)\phi_T(r,z)-\left[z^2+(1-z)^2\right]m_cK_1(m_cr)\partial_r\phi_T(r,z)\right\}\label{Phiphoton}
\end{equation}
with \cite{Kowalski:2003hm,Marquet:2007qa}
\begin{equation}
\phi_T(r,z)=N_Tz(1-z)\exp\left[-\frac{r^2}{2R_T^2}\right]\label{phiT}
\end{equation}
and where $N_T=1.23$, $R_T^2=6.5$ GeV$^{-2}$ \cite{Marquet:2007qa}.

Except for a factor of $z(1-z)$ in (\ref{phiT}) our notation, and choice of $J/\psi$ wave function exactly matches that of Ref.~\cite{Kowalski:2006hc}. Because (\ref{factxsec}) is a collinear factorized expression the gluon projectile on the right hand side of (\ref{factxsec}) is on-shell and so only transverse polarizations appear. We have taken the photon in (\ref{xsecphoton}) also on-shell so that the relationship between the photon and gluon induced processes will involve only a normalization change in (\ref{Phiphoton}) and a change of the $1-S$ factor in (\ref{Aphoton}).

We can get (\ref{xsecphoton}) in a more convenient form by using (\ref{Aphoton}) and integrating over $\Delta$. Thus
\begin{equation}
\frac{d\sigma_{\gamma A\to J/\psi A'}}{d^2b}=\int_0^1dz\int\frac{d^2r}{4\pi}\Phi_\gamma(\boldsymbol r,z)\int_0^1dz'\int\frac{d^2r'}{4\pi}\Phi^*_\gamma(\boldsymbol{r'},z')\left[1-S^*(r')\right]\left[1-S(r)\right]\label{xsecphA}
\end{equation}
where we have suppressed the energy and impact parameter dependence in the $1-S$ factors in (\ref{xsecphA}). The $S$ factors are given, in the McLerran-Venugopalan model \cite{MV}, by
\begin{equation}
S(r)=\exp\left[-\frac{C_F}{N_c}\frac{Q_s^2}{4}r^2\right]\simeq\exp\left[-\frac{1}{8}Q_s^2r^2\right]\label{mvdipole}
\end{equation}
and the cross section in (\ref{xsecphA}) allows nuclear breakup but is elastic at the dipole-nucleon scattering level. $Q_s$ in (\ref{mvdipole}) is the gluon saturation momentum with impact parameter dependence again suppressed.

\begin{figure}
\begin{center}
\includegraphics[width=16cm]{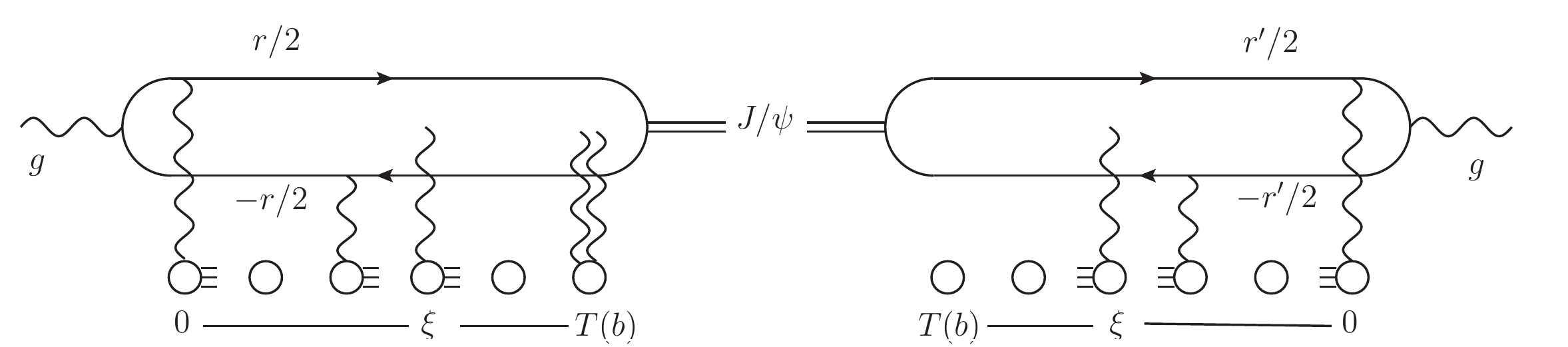}
\end{center}
\caption{Smaple diagram contributing to the $gA\to J/\psi$ process. The point of the last inelastic interaction is signaled explicitly at the longitudinal coordinate $\xi$.}
\label{scatpic}
\end{figure}

The main change necessary to convert (\ref{xsecphA}) to a cross section for $gA\to J/\psi X$ is the way the $c\bar{c}$ dipole scatters off nucleons in the nucleus. In (\ref{xsecphA}) the scatterings are purely elastic, and such scatterings are dominant in the large-$N_c$ limit because the quantum numbers of the $\gamma$ and the $J/\psi$ are the same. In $gA$ collisions the $c\bar{c}$ pair emerging from the gluon is in the adjoint color representation. The $c\bar{c}$ forming the $J/\psi$ is, of course, a color singlet. In the large-$N_c$ approximation there is a particular dipole-nucleon inelastic collision which converts the adjoint representation to a color singlet. This inelastic interaction is at the longitudinal coordinate $\xi$, starting from the front of the nucleus, in Fig. \ref{scatpic}. Later interactions, occurring after the $c\bar{c}$ pair is in a singlet state, are purely elastic in order to keep the singlet intact. Earlier interactions, occurring while the $c\bar{c}$ is in the adjoint representation, may be either elastic, occurring off a \emph{single} $c$ or $\bar{c}$ in the amplitude or complex conjugate amplitude, or inelastic involving the $c$ in both the amplitude and complex conjugate amplitude or involving the $\bar{c}$ in both the amplitude and complex conjugate amplitude. Sample interactions are illustrated in Fig. \ref{scatpic}.

The interaction at $\xi$ gives the factor
\begin{equation}
\frac{Q_s^2\;\boldsymbol r\cdot\boldsymbol{r'}}{4T(b)}d\xi.\label{lastinel}
\end{equation}
The interactions occurring before $\xi$ give the factor
\begin{equation}
e^{-\frac{1}{16}Q_s^2(\boldsymbol r-\boldsymbol{r'})^2(\xi/T(b))}\label{intbefore}
\end{equation}
while those occurring after give
\begin{equation}
e^{-\frac{1}{8}Q_s^2(r^2+r^{\prime2})(1-\xi/T(b))}.\label{intafter}
\end{equation}
In going from $\gamma A\to J/\psi A'$ to $gA\to J/\psi X$ the $\left[1-S^*(r')\right]\left[1-S(r)\right]$ factor in (\ref{xsecphA}) gets replaced by the product of the factors in (\ref{lastinel})-(\ref{intafter}).

\begin{figure}
\begin{center}
\includegraphics[width=16cm]{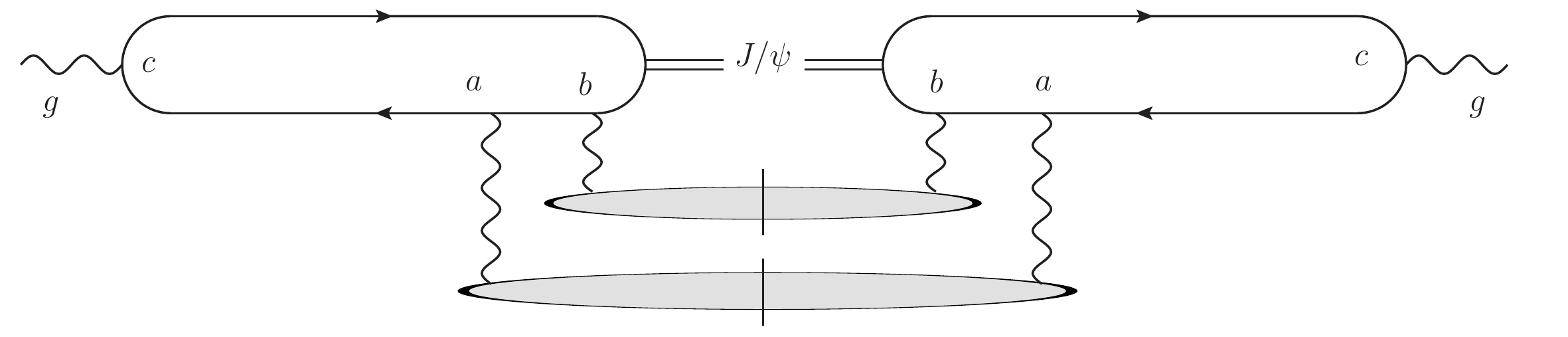}
\end{center}
\caption{Lowest order process in gluon induced $J/\psi$ production. Color indices are indicated explicitly.}
\label{lorder}
\end{figure}

In addition there is a color factor. In $\gamma$ induced $J/\psi$ production there is a factor of $N_c$ in the amplitude and a factor of $N_c$ in the complex conjugate amplitude. This is the factor of $N_c$ explicit in (\ref{Phiphoton}) coming from a sum over the colors of the $c$, and $\bar{c}$, making up the $J/\psi$. To find the color factors in the gluon induced process it is necessary to evaluate the lowest order process shown in Fig. \ref{lorder}. The color factors for this process are, as already found in \cite{Kharzeev:2008cv},
\begin{equation}
\frac{1}{(N_c^2-1)^3}\text{Tr}\left(t^ct^at^b\right)\text{Tr}\left(t^ct^bt^a\right)\simeq\left(\frac{C_F}{N_c^2-1}\right)^2\frac{1}{2N_c}\label{cfactor}
\end{equation}
where we have used the large-$N_c$ limit in the right hand side of (\ref{cfactor}). The $\left(C_F/(N_c^2-1)\right)^2$ factors go into making up part of the two factors of $Q_s^2$ that come from the graphs. Explicit calculation confirms that the remaining factor, after taking out the factor in (\ref{lastinel}) and the factor linear in $Q_s^2$ when expanding (\ref{intbefore}) is just the factor $1/2N_c$ on the right hand side of (\ref{cfactor}).

Putting all this together gives
\begin{align}
\frac{d\sigma_{gA\to J/\psi X}}{d^2b}=&\;\int_0^1dz\int\frac{d^2r}{4\pi}\Phi(\boldsymbol r,z)\int_0^1dz'\int\frac{d^2r'}{4\pi}\Phi^*(\boldsymbol{r'},z')\nonumber\\
&\times\int_0^{T(b)}d\xi\;\frac{\boldsymbol r\cdot\boldsymbol{r'}Q_s^2}{4T(b)}\exp\left\{-\frac{1}{16}Q_s^2(\boldsymbol r-\boldsymbol{r'})^2\frac{\xi}{T(b)}-\frac{1}{8}Q_s^2(r^2+r^{\prime2})\left(1-\frac{\xi}{T(b)}\right)\right\}
\end{align}
with
\begin{equation}
\Phi(\boldsymbol r,z)=\left[\frac{2}{3}eN_c\right]^{-1}\frac{g}{\sqrt{2N_c}}\Phi_\gamma(\boldsymbol r,z)\label{Phig}
\end{equation}
where, finally, in (\ref{Phig}) we have introduced the replacement $\frac{2}{3}e\to g$. Doing the integral over $\xi$ and using (\ref{factxsec}) we get
\begin{align}\label{pa-m}
\frac{d\sigma_{pA\to J/\psi X}}{dyd^2b}=&\;x_1G(x_1,m_c^2)\int_0^1dz\int\frac{d^2r}{4\pi}\Phi(\boldsymbol r,z)\int_0^1dz'\int\frac{d^2r'}{4\pi}\Phi^*(\boldsymbol{r'},z')\nonumber\\
&\times\frac{4\boldsymbol r\cdot\boldsymbol{r'}}{(\boldsymbol r+\boldsymbol{r'})^2}\left(e^{-\frac{Q_s^2}{16}(\boldsymbol r-\boldsymbol{r'})^2}-e^{-\frac{Q_s^2}{8}(r^2+r^{\prime2})}\right).
\end{align}

\section{$\jpsi$ cross section in AA collisions}\label{sec:AA}

Generalization of the result of the previous section to nucleus-nucleus collisions is achieved by letting the initial gluon be emitted from either nucleus and taking into account $c\bar c$ dipole scattering in both nuclei. The scattering amplitudes and the saturation scales for the two nuclei depend on their respective impact parameters $\b b_1$ and $\b b_2$. To make our notations more compact we will not  indicate the impact parameter dependence explicitly. Introducing  the relative impact parameter
 $\b B= \b b_1-\b b_2$  and  using the relation 
\beq \label{aa1}
\frac{\as \pi^2}{4C_F} x_1G(x_1,a^2)= \int d^2b_1\,\frac{Q^2_{s1}}{8}
\eeq
we can write the cross section as 
\begin{align}\label{ab-2}
\frac{d\sigma_{A_1A_2\to J/\psi X}}{dy\,d^2b\, d^2B}= \int_0^1 dz\int \frac{d^2\b r}{4\pi}\int_0^1dz'\int\frac{d^2\b r'}{4\pi}\,\Phi(\b r,z)\, \Phi^*_{\lambda\lambda'}(\b r',z')\,2T_{A_1A_2\to JX}(\b r, \b r')\,,
\end{align}
where 
\begin{align}\label{aa-m}
T_{A_1A_2\to JX}(\b r, \b r')= \frac{C_F}{2\as\pi^2}\frac{Q_{s1}^2Q_{s2}^2}{Q_{s1}^2+Q_{s2}^2}\frac{4\b r\cdot \b r'}{(\b r+\b r')^2}\left( 
e^{-\frac{1}{16}(Q_{s1}^2+Q_{s2}^2)(\b r-\b r')^2}-
e^{-\frac{1}{8}(Q_{s1}^2+Q_{s2}^2)(r^2+r'^2)}\right)
\end{align}
Expanding \eq{aa-m} at small $Q_{s1}^2$ we recover Eq. \eq{pa-m}. 

The first few terms in the expansion of \eq{aa-m} in nuclear density read
\begin{align}\label{aa-exp}
T_{A_1A_2\to JX}(\b r, \b r')\approx \frac{C_F}{2\as\pi^2} Q_{s1}^2Q_{s2}^2\,4\b r\cdot \b r'\,\left( \frac{1}{16} - \frac{1}{512}(Q_{s1}^2+Q_{s2}^2)(3r^2+3r'^2-2\b r\cdot \b r')\right)
\end{align}
Averaging over the relative angle between $\b r$ and $\b r'$ yields
\begin{align}\label{aa-exp2}
\aver{T_{A_1A_2\to JX}(\b r, \b r')}\approx \frac{C_F}{\as\pi^2}\frac{r^2r'^2}{16^2}\left( Q_{s1}^2Q_{s2}^4+ Q_{s1}^4Q_{s2}^2\right)
\end{align}
This is the leading contribution to the $\jpsi$ production; it is easily seen that it breaks the factorization.
We believe that \eq{aa-m} is a reasonable starting point for phenomenology of $\jpsi$ production in $AA$ collisions. Nevertheless a better theoretical understanding of the $AA$ production amplitude $T_{A_1A_2\to
J/\psi X}$ is desirable.

\section{Rapidity and energy dependence}\label{sec:evol}

Eqs.~\eq{ab-2},\eq{aa-m}  can be readily generalized to include quantum evolution effects. To that end we recall that the initial condition for the BK \cite{Balitsky:1995ub,Kovchegov:1999yj} evolution equation is given by the Glauber--Mueller formula for the forward dipole--nucleus quark dipole elastic scattering amplitude \cite{Mueller:1989st}
\beql{init.cond}
N_F(\b r, \b b, y_0)= 1-e^{-\frac{1}{8}\b r^2Q_{s}^2(y_0)}\,,
\eeq
where subscript $F$ indicates the fundamental representation. Evolution of the gluon dipole scattering amplitude (adjoint representation) obeys the equation 
\beql{init.cond.G}
N_A(\b r, \b b, y)= 2N_F(\b r, \b b, y)-N_F^2(\b r, \b b, y)
\eeq
and its initial condition is 
\beql{init.cond.a}
N_A(\b r, \b b, y_0)= 1-e^{-\frac{1}{4}\b r^2Q_{s}^2(y_0)}\,,
\eeq
Accordingly, we can incorporate evolution effects in \eq{aa-m} by the following replacements \cite{Kovchegov:2000hz}
\begin{align}
e^{-\frac{1}{8}Q_s^2r^2}&\to 1-N_F(\b r, \b b, y)\\
e^{-\frac{1}{16}Q_s^2 r^2} &\to  1-N_A(\b r/2,\b b, y)
\end{align}
Omitting the impact parameter dependence as before, we thus obtain 
\begin{align}\label{aa-m-ev}
T_{A_1A_2\to JX}(\b r, \b r')= \frac{8N_c}{\as\pi^2}\frac{Q_{s1}^2Q_{s2}^2}{Q_{s1}^2+Q_{s2}^2}\frac{4\b r\cdot \b r'}{(\b r+\b r')^2} \left\{ 
\left[1-N_A^{(1)}((\b r-\b r')/2,  y)\right] \left[1-N_A^{(2)}((\b r-\b r')/2, -y)\right]  \right. &\nonumber\\
\left. -\left[1-N_F^{(1)}(\b r, y)\right]\left[1-N_F^{(1)}(\b r', y)\right]
\left[1-N_F^{(2)}(\b r, -y)\right]\left[1-N_F^{(2)}(\b r', -y)\right]\right\} & 
\end{align}

\section{Numerical calculations}\label{sec:numeric}

The experimental data is expressed in terms of the nuclear modification factor (NMF) defined as
\beql{nmf}
R_{A_1A_2}= \frac{\int_\mathcal{S}\, d^2b\frac{d\sigma_{A_1A_2\to J/\psi X}}{dy\, d^2b}}
{A_1\,A_2\,\frac{d\sigma_{pp\to J/\psi X}}{dy}}\,.
\eeq
where $\mathcal{S}$ stands for the overall area of two colliding nuclei. Since the mechanism of $\jpsi$ production in $pp$ collisions remains elusive, we follow our approach in the previous publications and approximate
\beql{prev}
\frac{d\sigma_{pp\to \jpsi X}}{dy}= C\, \frac{d\sigma_{AA\to \jpsi X}}{dy}\bigg|_{A=1}
\eeq
with $C=\text{const}$. We fix the constant to provide the best description of the $pp$ and $dA$ data. It is reassuring that the numerical calculations described in the next section indicate that $C$ is close to unity.
 
\begin{figure}[ht]
\begin{tabular}{cc}
    \includegraphics[width=7.5cm]{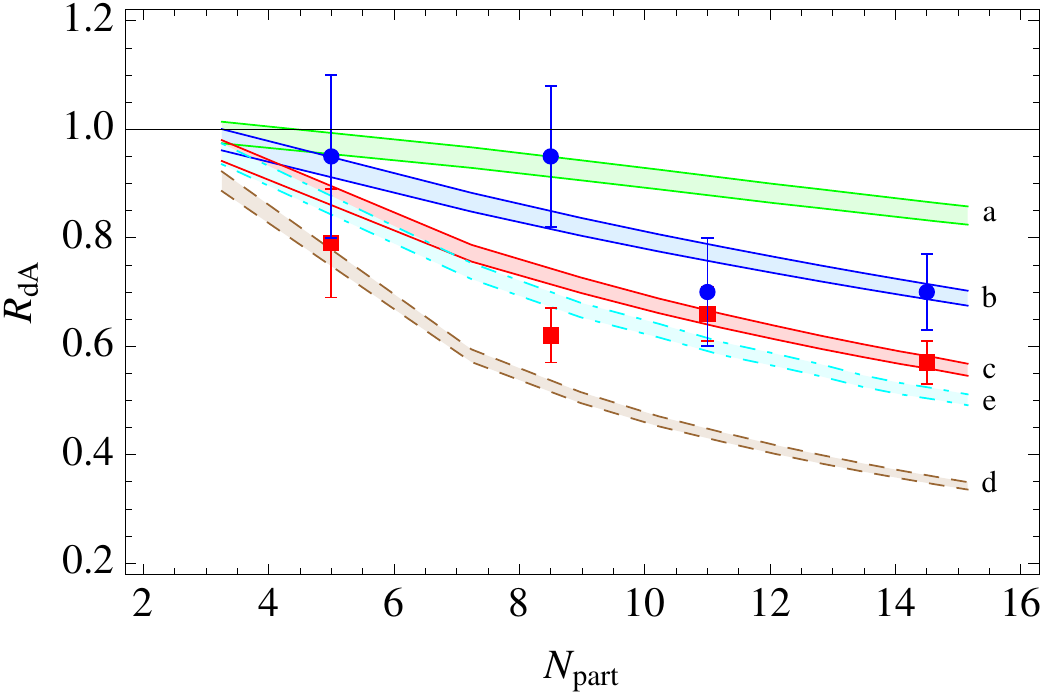} &
     \includegraphics[width=7.5cm]{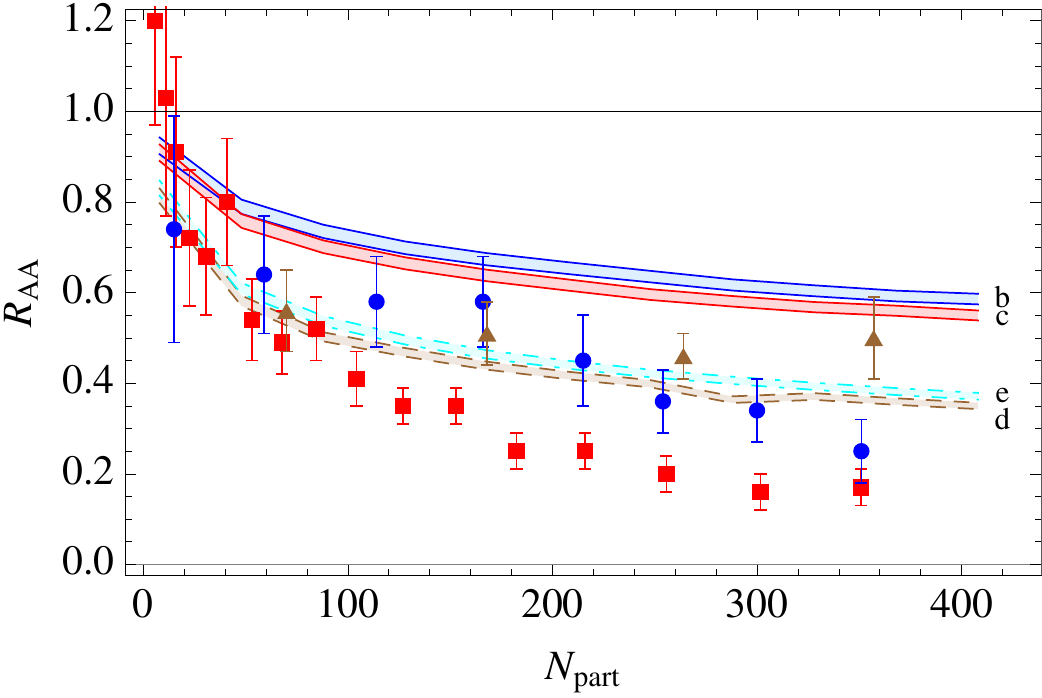} \\
     (a) & (b)
  \end{tabular}
  \caption{ 
 Nuclear modification factor vs $N_\mathrm{part}$ in (a) $dAu$ and (b) $AA$ collisions using the DHJ model
 \cite{Dumitru:2005kb}.  Band `a' (green) represents  rapidity $y=-1.7$ at $\sqrt{s}=200$~GeV, `b' (blue): $y=0$, $\sqrt{s}=200$~GeV, `c' (red):  $y=1.7$, $\sqrt{s}=200$~GeV, `d' (brown): $y=3.25$, $\sqrt{s}=2.76$~TeV, `e' (cyan): $y=0$, $\sqrt{s}=5.5$~TeV.  $m=1.5$~GeV, $C=1$. Experimental data \cite{Adler:2005ph,Adare:2006ns,Adare:2011yf,Pillot:2011zg} is represented by (blue) circles in `b', by (red)  squares in `c' and by (brown) triangles in `d'.  (Color online).}
  \label{fig:dhj}
\end{figure}

\begin{figure}[ht]
\begin{tabular}{cc}
     \includegraphics[width=7.5cm]{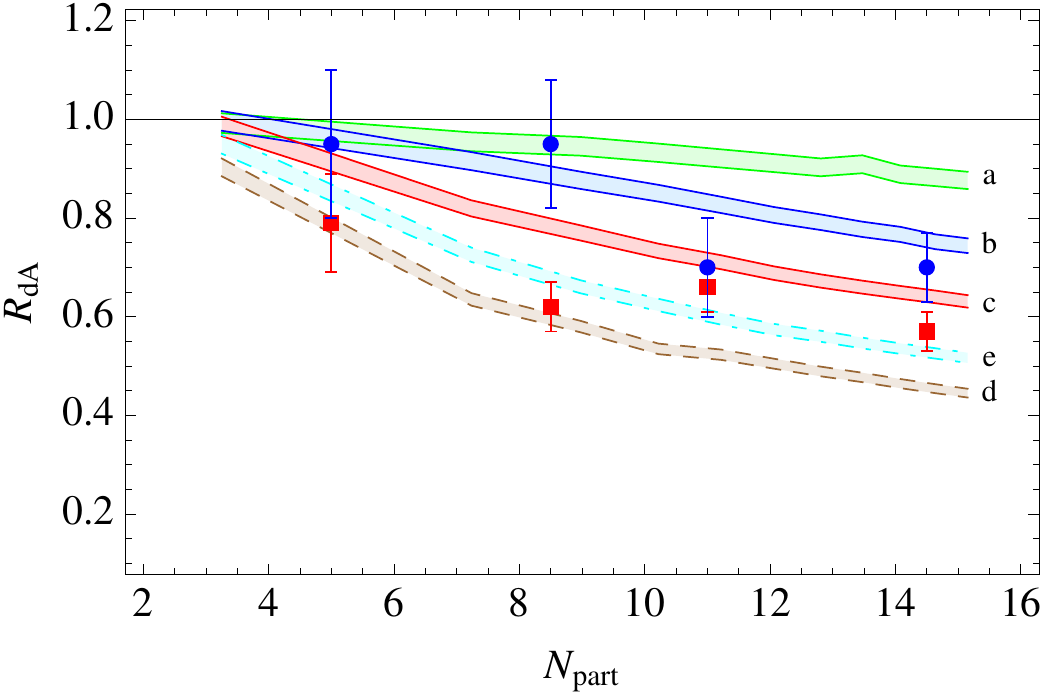} &
    \includegraphics[width=7.5cm]{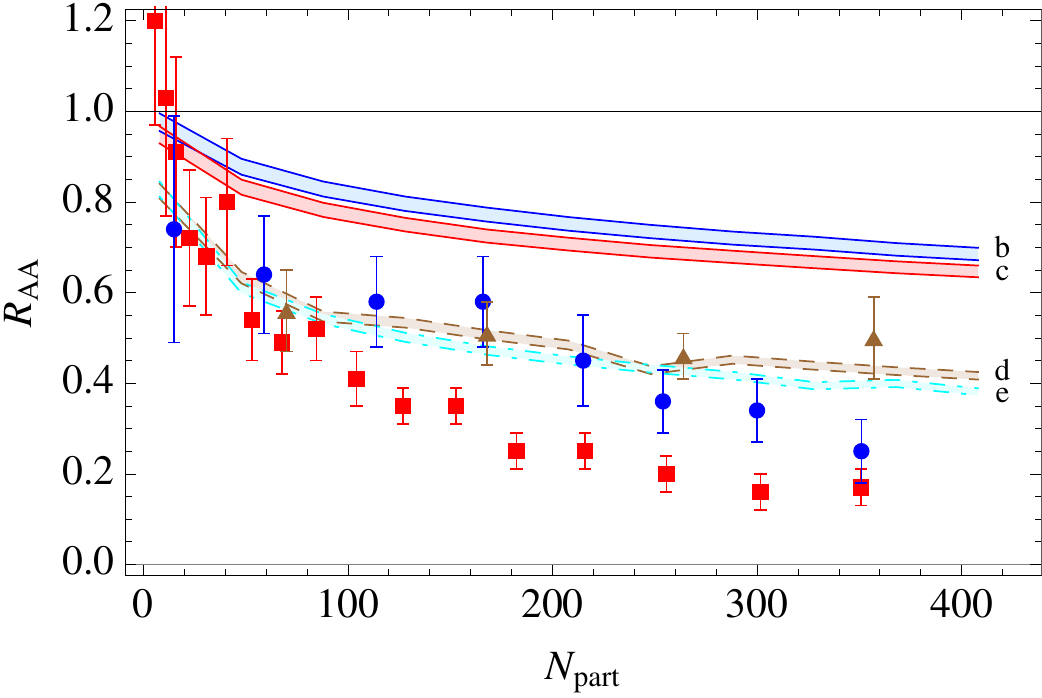} \\
    (a) & (b)
  \end{tabular}
 \caption{  Same as \fig{fig:dhj} using the bCGC model \cite{Kowalski:2006hc}. }
\label{fig:kmw}
\end{figure}

The results of our calculations are exhibited in \fig{fig:dhj} and   \fig{fig:kmw}; we have used two different models for the dipole scattering amplitude:  DHJ \cite{Dumitru:2005kb} and bCGC \cite{Kowalski:2006hc} models   (see Appendix~\ref{appA} for the description of these models). Comparison of the results of the two models gives an idea about the model dependence of the numerical results. We observe a reasonable agreement with the experimental data on $J/\psi$ production in $dA$ collisions. 

Concerning the $\jpsi$ production in $AA$ collisions all models underestimate the suppression at RHIC both at mid-rapidity and in the forward rapidity. Moreover, it appears that the gluon saturation effects on NMF show very little rapidity dependence at RHIC which contradicts the experimental data.
We also find that there is almost no change between the NMF at  LHC $\sqrt{s}= 2.76$~TeV and $5.5$~TeV. We note that  our calculation overestimates the NMF at $\sqrt{s}= 2.76$~TeV.

\section{Discussion and conclusions} \label{sec:disc}

Our calculations indicate that the nuclear modification of $\jpsi$ production  in $dA$ collisions at RHIC is dominated by the cold nuclear matter effects. It would be important to study $\jpsi$ production  in $pA$ collisions at LHC; \fig{fig:dhj} and \fig{fig:kmw} provide our predictions. In contrast, the cold nuclear matter effects alone cannot provide neither quantitative nor even a qualitative description of the $AA$ data. Additional mechanisms beyond  the initial state effects are required to explain the experimental data. It is remarkable that at RHIC these additional mechanisms must provide extra suppression of the NMF, perhaps via the Matsui-Satz color screening mechanism
 \cite{Matsui:1986dk} or the gluon-induced dissociation \cite{Shuryak,Kharzeev:1994pz}, whereas at LHC they must produce enhancement.   
 
Our successful description of the $\jpsi$ NMF in $pA$ collisions with the normalization factor $C=1$ in \eq{prev} may be an evidence that the $\jpsi$ production mechanism in $pp$ collisions is similar to that in $pA$ implying that it is perhaps dominated by the higher twist effects. 

To summarize, we derived the formulae for the cross sections of $\jpsi$ production in $pA$ and $AA$ collisions taking into account the gluon saturation/color glass condensate effects. Our numerical results provide an estimate of the color nuclear matter effects on  $\jpsi$ production in heavy-ion collisions.

\acknowledgments
The work of  D.K.\ was supported in part by the U.S.\ Department of Energy under Contracts No. DE-AC02-98CH10886 and DE-FG-88ER41723.
K.T.\ was supported in part by the U.S.\ Department of Energy under Grant No.\ DE-FG02-87ER40371. This research of E.L.\ was supported in part by the Fondecyt (Chile) grant 1100648.

\appendix
\section{Models of the dipole scattering amplitude}\label{appA}

We performed  numerical calculations  using two models of the dipole scattering amplitude: DHJ \cite{Dumitru:2005kb} and bCGC \cite{Kowalski:2006hc} models.  The DHJ model is an improvement of the  KKT model\cite{Kharzeev:2004yx,Tuchin:2007pf} that takes into account the change in the anomalous dimension of the gluon distribution function due to the presence of the saturation boundary \cite{Mueller:2002zm} and takes into account some higher order effect. It successfully describes the single inclusive hadron production in $dA$ collisions in the relevant kinematic region.  In this model, the dipole scattering amplitude is parameterized as follows
\beq\label{modN}
N_A(\b r,0, y)=1-\exp\left\{ -\frac{1}{4}\left(r^2 Q_s^2\right)^{\gamma}\right\}\,.
\eeq
 The \emph{gluon} saturation scale  is given by
\beq\label{satt}
Q_s^2=\Lambda^2\, A^{1/3}\, e^{\lambda y}=0.13\,\mathrm{GeV}^2\,e^{\lambda y}\,N_\mathrm{coll}\,.
\eeq
where the parameters $\Lambda=0.6$ GeV and $\lambda=0.3$ are fixed by DIS data \cite{MOD}.
\begin{align}
\gamma= \gamma_s + (1-\gamma_s)\frac{\ln(m^2/Q_s^2)}{\lambda Y+ \ln( m^2/Q_s^2)+d \sqrt{Y} }
\end{align}
 where $\gamma_s= 0.628$ is implied by theoretical arguments \cite{Mueller:2002zm} and $d=1.2$ is fixed by fitting to the hadron production data in $dA$ collisions at RHIC.  $Y= \ln(1/x)$, with $x= me^{-y}/\sqrt{s}$. The quark dipole scattering amplitude is given by 
 \begin{align}
 N_F(\b r, 0, y)= 1-\sqrt{1-N_A(\b r, 0, y)}
 \end{align}
 which follows from \eq{init.cond.G}.

We used  the bCGC model  \cite{Kowalski:2006hc} with a modification: we treat the nuclei and proton profiles as step-functions; the saturation scales are assumed to scale with $A$ as $Q_s^2\propto A^{1/3}$. The advantage of this model -- besides its compliance with the known analytical approximations to the BK equation \cite{Iancu:2002tr} -- is that its parameters are fitted to the low $x$ DIS data. The explicit form of the scattering amplitude $N$   is given by
\beql{kmw}
N_F(\b r, 0, y)= \,\left\{
\begin{array}{cc}
\mathcal{N}_0\left( \frac{r^2\Q_s^2}{4}\right)^\gamma\,,&\quad r\Q_s\le 2;\\
1-\exp[-a\ln^2(br\Q_s)]\,,&\quad r\Q_s\ge 2\,,
\end{array}\right.
\eeq
where $\Q_s^2$ is the the \emph{quark} saturation scale related to the \emph{gluon} saturation scale $Q_s^2$ -- which we have called simply the `saturation scale' throughout the paper -- by $\Q_s^2= (4/9)Q_s^2$. Its functional form is
\beql{sat.scale}
\Q_s^2= A^{1/3} x_0^\lambda\, e^{\lambda y}\,s^{\lambda/2}\,\mathrm{GeV}^2\,,
\eeq
where $s$ is the square of the center-of-mass energy and $y$ is rapidity with respect to the central rapidity. The anomalous dimension is 
\beql{anom.dim}
\gamma = \gamma_s+\frac{1}{c\, \lambda \,(\ln\surd s+y)}\ln \left( \frac{2}{r\Q_s}\right)\,.
\eeq
Parameters $\gamma_s=0.628$ and $c=9.9$ follow from the BFKL dynamics \cite{Iancu:2002tr}, while $\mathcal{N}_0=0.7$ and $\lambda=0.28$ are fitted to the DIS data. Constants $a$ and $b$ are uniquely fixed from by the requirement of continuity of the amplitude and its first derivative.



\begin{thebibliography}{99}

\bibitem{Kharzeev:2005zr}
  D.~Kharzeev and K.~Tuchin,
  Nucl.\ Phys.\  A {\bf 770}, 40 (2006)
  [arXiv:hep-ph/0510358].

\bibitem{Kharzeev:2008cv}
  D.~Kharzeev, E.~Levin, M.~Nardi and K.~Tuchin,
  Nucl.\ Phys.\  A {\bf 826}, 230 (2009)
  [arXiv:0809.2933 [hep-ph]].

\bibitem{Kharzeev:2008nw}
  D.~Kharzeev, E.~Levin, M.~Nardi and K.~Tuchin,
  Phys.\ Rev.\ Lett.\  {\bf 102}, 152301 (2009)
  [arXiv:0808.2954 [hep-ph]].


\bibitem{MV}
L.~D.~McLerran and R.~Venugopalan,
Phys.\ Rev.\ D {\bf 49}, 2233 (1994),
Phys.\ Rev.\ D {\bf 49}, 3352 (1994),
   
\bibitem{Kovchegov:1999yj}
  Y.~V.~Kovchegov,
  Phys.\ Rev.\  D {\bf 60}, 034008 (1999)
  [arXiv:hep-ph/9901281].
 
\bibitem{Kovchegov:1996ty}
  Y.~V.~Kovchegov,
  Phys.\ Rev.\  D {\bf 54} (1996) 5463
  [arXiv:hep-ph/9605446].

\bibitem{Kowalski:2003hm}
  H.~Kowalski and D.~Teaney,
  Phys.\ Rev.\  D {\bf 68}, 114005 (2003)
  [arXiv:hep-ph/0304189].

\bibitem{Marquet:2007qa}
  C.~Marquet, R.~B.~Peschanski and G.~Soyez,
  Phys.\ Rev.\  D {\bf 76}, 034011 (2007)
  [arXiv:hep-ph/0702171].

\bibitem{Kowalski:2006hc}
  H.~Kowalski, L.~Motyka and G.~Watt,
  Phys.\ Rev.\  D {\bf 74}, 074016 (2006)
  [arXiv:hep-ph/0606272].

\bibitem{Balitsky:1995ub}
I.~Balitsky,
Nucl.\ Phys.\ B {\bf 463}, 99 (1996)
[arXiv:hep-ph/9509348];

\bibitem{Mueller:1989st}
  A.~H.~Mueller,
  Nucl.\ Phys.\  B {\bf 335}, 115 (1990).

\bibitem{Kovchegov:2000hz}
  Y.~V.~Kovchegov,
  Nucl.\ Phys.\  A {\bf 692}, 557 (2001)
  [arXiv:hep-ph/0011252].

\bibitem{Dumitru:2005kb}
  A.~Dumitru, A.~Hayashigaki, J.~Jalilian-Marian,
  Nucl.\ Phys.\  {\bf A770}, 57-70 (2006).
  [hep-ph/0512129].
  
\bibitem{Kharzeev:2004yx}
  D.~Kharzeev, Y.~V.~Kovchegov and K.~Tuchin,
  Phys.\ Lett.\  B {\bf 599}, 23 (2004)
  [arXiv:hep-ph/0405045].

\bibitem{Mueller:2002zm}
  A.~H.~Mueller, D.~N.~Triantafyllopoulos,
  Nucl.\ Phys.\  {\bf B640}, 331-350 (2002).
  [hep-ph/0205167].









  
\bibitem{Adler:2005ph}
  S.~S.~Adler {\it et al.}  [PHENIX Collaboration],
  Phys.\ Rev.\ Lett.\  {\bf 96}, 012304 (2006)
  [arXiv:nucl-ex/0507032].



\bibitem{Adare:2006ns}
  A.~Adare {\it et al.}  [PHENIX Collaboration],
  Phys.\ Rev.\ Lett.\  {\bf 98}, 232301 (2007)
  [arXiv:nucl-ex/0611020].
  
 
\bibitem{Adare:2011yf}
  A.~Adare {\it et al.},
  arXiv:1103.6269 [nucl-ex].
  
\bibitem{Pillot:2011zg}
  P.~Pillot, f.~t.~A.~Collaboration,
   [arXiv:1108.3795 [hep-ex]].
 
  
\bibitem{Matsui:1986dk}
  T.~Matsui and H.~Satz,
  Phys.\ Lett.\  B {\bf 178}, 416 (1986).
 
\bibitem{Tuchin:2007pf}
  K.~Tuchin,
  Nucl.\ Phys.\  A {\bf 798}, 61 (2008)
  [arXiv:0705.2193 [hep-ph]].
  
  \bibitem{MOD}
   K.~J.~Golec-Biernat and M.~Wusthoff,
  Phys.\ Rev.\  D {\bf 59} (1998) 014017
  [arXiv:hep-ph/9807513],
  Phys.\ Rev.\  D {\bf 60} (1999) 114023
  [arXiv:hep-ph/9903358].


  
\bibitem{Iancu:2002tr}
  E.~Iancu, K.~Itakura and L.~McLerran,
  Nucl.\ Phys.\  A {\bf 708}, 327 (2002)
  [arXiv:hep-ph/0203137].

\bibitem{Shuryak}
E.V. Shuryak, Sov. J. Nucl. Phys {\bf 28} (1978) 408.

\bibitem{Kharzeev:1994pz}
  D.~Kharzeev, H.~Satz,
  Phys.\ Lett.\  {\bf B334}, 155-162 (1994).
  [hep-ph/9405414].

\end{thebibliography}
\end{document}